\documentstyle[12pt]{article}
\begin{document}
\setlength{\baselineskip}{0.30in}

\newcommand{\beq}{\begin{equation}}
\newcommand{\eeq}{\end{equation}}

\newcommand{\bi}{\bibitem}
\def\mpl{m_{Pl}}

%{\hbox to\hsize{Feb. 1994 \hfill UM-AC-94-00}}
%{\hbox to\hsize{March 29, 1994 \hfill Draft-Einhorn 10}}
\begin{flushright}
UM - TH - 94 - 11\\
\today\\
gr-qc/9403056
\end{flushright}

\begin{center}
\vglue .06in
{\Large \bf {On Infrared Effects in de~Sitter Background}}\\[.5in]

{\bf A. D. Dolgov,\footnote{Permanent address: ITEP, 113259, Moscow,
Russia.} M. B. Einhorn, and V. I. Zakharov}
\\[.05in]
{\it{The Randall Laboratory of Physics\\
University of Michigan\\
Ann Arbor, MI 48109}}\\[.15in]
\end{center}
%{Abstract}\\[-.1in]
\begin{abstract}
\begin{quotation}
We have estimated higher order quantum gravity corrections to de~Sitter
spacetime. Our results suggest that, while the classical spacetime metric may
be distorted by the graviton self-interactions, the corrections are
relatively weaker than previously thought, possibly growing like a power
rather than exponentially in time.
\end{quotation}
\end{abstract}

\newpage

As is very well known Einstein field equations permit an addition of an extra
term proportional to the metric tensor, $\bar g_{\mu\nu}$:
\beq{
\bar R_{\mu\nu} -{1\over 2} \bar g_{\mu\nu} \bar R =
8\pi G T_{\mu\nu} + \Lambda \bar g_{\mu\nu}
}\label{eineq}
\eeq
Here $T_{\mu\nu}$ is the energy-momentum tensor of matter and $\Lambda$ is the
so called cosmological constant. (We put an bar over all quantities here to
distinguish them from the rescaled ones, see Eqs.~(\ref{psi},\ref{action})).
The last term in Eq.~(\ref{eineq}) may be interpreted as the vacuum
energy-momentum tensor. Any natural estimate gives $\Lambda$ or
$\rho_{vac}=\Lambda/8\pi G$ much larger (by 50-100 orders of magnitude) than
the observed upper bound in the present day Universe. This mysterious
discrepancy is known as the cosmological constant problem and presents one of
the most interesting challenges in the modern physics (for a review see
Refs.~\cite{sw,ad}).

de~Sitter spacetime is a solution of the Einstein equations with a dominant
cosmological term. Written in the Robertson-Walker form for the special case
of the spatially flat section the metric takes the form
\beq{
ds^2=dt^2-a^2(t) d\vec{r}\,^2
}\label{dsphys}
\eeq
where $a(t)=\exp (Ht)$ with $H=\sqrt {\Lambda/3}$. The assumption of spatial
flatness is not essential and the results obtained below are true also for
open and closed geometries.  It is convenient to rewrite the metric in terms
of conformal coordinates where, up to an overall scale factor, it has the
Minkowski form:
\beq{
ds^2= a^2(\tau) (d\tau^2 - d\vec {r}\,^2)
}\label{dsconf}
\eeq
The conformal time $\tau $ is related to the physical one as
$d\tau =-\exp (-Ht)dt$ and $a(\tau) =-1/H\tau$.  Note that when $t$ tends to
future infinity, $\tau$ tends to $0-$.  The de~Sitter metric possesses the
same degree of high symmetry as Minkowski one and for this reason, it is the
simplest
non-trivial curved background for a quantum field theory.  Despite its
simplicity, it has several very interesting properties and in particular
generates an infrared instability of a massless scalar field $\phi$ minimally
coupled to gravity. It was shown \cite{fp, bd} that the vacuum expectation
value (VEV) $\langle \phi^2 \rangle$ is singular at the zero mass limit,
$\langle\phi^2\rangle \sim H^4/m^2$.  With the advent of the inflationary
scenario,
this phenomenon was rediscovered in a number of papers \cite{adl,aas,vf} where
it was argued that $\langle \phi^2 \rangle$ rises linearly with time in the
zero mass limit.  This quantity however does not have a direct physical
meaning and, in particular, the contribution of the field $\phi$ to the energy
density does not rise with time.  Nevertheless, this infrared instability
afflicts the scalar field propagator and needs to be resolved to make a
sensible quantum theory.  Note in passing that this instability is a
result of broken conformal invariance.  To ensure the latter one needs the
coupling to the curvature scalar, $R\phi^2 /6$, which gives rise to
an infrared cut-off. In fact, any nonzero mass $m_\phi$ or coupling
$\xi R\phi^2$ with a nonnegative coefficient is sufficient for infrared
stability.  In any case, this infrared divergence is a sign that the true
vacuum state is not de~Sitter invariant.

Conformal invariance is also broken for gravitons \cite{lpg} and,
for this reason, there should be infrared instability of the de~Sitter
vacuum due to quantum gravity effects. If this is indeed the case, the
solution of the long standing cosmological constant problem may be found in
this direction. One-loop graviton quantum corrections to the de~Sitter metric
were considered in this connection in Ref.~\cite{lf} where it was found that
they are time independent.  However, it was suggested that higher loops may
show evidence of this infrared instability. Indeed, it was claimed recently in
several papers
\cite{tw1,tw2,tw3} that higher loop effects are much stronger, giving
corrections that rise exponentially with physical time $t$ or as a power of
$1/\tau$ with conformal time.  This is a very exciting result and, if
confirmed, would mean that de~Sitter space cannot exist indefinitely, opening
a beautiful way for the solution of the cosmological constant problem in the
framework of the normal quantum gravity without any drastic assumptions.

Here we have reconsidered results of papers \cite{tw1,tw2,tw3} using a
different formalism and have found, unfortunately, that such a strong
instability does not set in.  The corrections at most behave as a
power\footnote{We understand that these authors now also do not get power law
singularity but only powers of $\log
\tau$.~(R. Woodard, private communication)} of $\log|\tau| \sim t$.

For the conformally flat background metric we introduce the quantum graviton
field $h_{\mu\nu}$ in the usual way
\beq{
\bar g_{\mu\nu}=a^2(\tau ) g_{\mu\nu} \equiv
a^2(\tau) (\eta_{\mu\nu} + h_{\mu\nu}).
}\label{psi}
\eeq
The issue is whether, taking into account loop corrections, the VEV
$\langle h_{\mu\nu}\rangle$ is non-zero and, in particular,
whether it is time-dependent.  This would suggest that the background
de~Sitter metric is not self-consistent, although it must then be shown that
this is a physical effect by calculating, for example, the curvature for the
modified metric.  Even then, it is important to ascertain whether the
inconsistency involves strong coupling or whether it is an essentially
negligible effect.  The Einstein action (with nonzero cosmological constant)
in terms of the new metric $g_{\mu\nu}$ can be rewritten as
\beq{
A={1\over \kappa^2} \int d^4x \sqrt{\bar g} (\bar R + 2\Lambda)
={1\over \kappa^2}\int d^4x a^2 \sqrt g \bigl(R+6{a_{,\alpha}a^{,\beta}
\over a^2} + 2\Lambda a^2 \bigr),
}\label{action}
\eeq
with $\kappa^2 \equiv 16\pi/\mpl^2$.
This implies the following equations of motion:
\beq{
R_{\mu\nu} -{1\over 2} g_{\mu\nu} R -\Lambda a^2g_{\mu\nu} +
{4a_{,\mu}a_{,\nu} \over a^2} -{2a_{;\mu\nu} \over a}
+g_{\mu\nu}
\left({2a_{;\alpha}^{;\alpha} \over a}-
{a_{,\alpha}a^{,\alpha} \over a^2}\right)
=0
}\label{eqmot}
\eeq
where tensor indices are raised and lowered with the new metric $g_{\mu\nu}$.
To zeroth order in $h_{\mu\nu}$, we get the usual equation for the scale
factor of the classical background metric, with solution $a(\tau)=-1/H\tau$
(see notation after Eq.~(\ref{dsconf}).)

It is of course well-known that this is a non-renormalizable quantum field
theory, but we regard the Einstein action as the first two terms in an
infinite series of local operators of increasing dimension.  This is an
effective field theory, presumed valid on scales below $\mpl$, and is
renormalizable in the sense that all divergences involving these vertices may
be absorbed in a renormalization of one of the infinite number of coefficients
of these local
operators.  While we have not written such terms explicitly, they should be
understood to be present.  In the following, we will deal exclusively with
renormalized couplings and operators, with the tacit understanding
that the counterterms are included in our interaction.  To consistently
quantize the theory, we must add a gauge-fixing term ${\cal L}_{gf}$ and the
corresponding Faddeev-Popov ghosts ${\cal L}_{FP}.$  We choose ${\cal
L}_{gf}=-{1\over2}F_\mu F_\nu\eta^{\mu\nu}$ as in Refs.~\cite{tw1,tw2}, with
\beq{
F_\mu=a(\tau)\bigl( h^\nu _{\mu,\nu} -{1\over 2} h_{,\mu} +
2\delta^0_\mu h^\nu _0 {a_{,\nu} \over a} \bigr)
}\label{gauge}
\eeq
The ghost Lagrangian ${\cal L}_{FP}$ may be found in Ref.~\cite{tw3}.
Here and subsequently when we consider perturbation theory in $h_{\mu\nu}$,
the indices are raised with the Minkowski tensor  $\eta_{\mu\nu}$. With these
gauge conditions, the linear part of the equation of motion for $h_{\mu\nu}$
has very simple form:
\beq{
h_{\mu\nu,\alpha}^\alpha - {2\over \tau} h_{\mu\nu,0}
+{2\over \tau^2} \left( \delta^0_\mu h_{0\nu} + \delta^0_{\nu}h_{0\mu}
\right) -{2\over \tau^2} \eta_{\mu\nu}h_{00} =0
}\label{first}
\eeq
One can easily verify that time components $h_{0\mu}$ ($\mu =0,1,2,3$)
are conformally invariant in the sense that the
\underline{rescaled} functions
$\chi_{0\mu} =a(\tau) h_{0\mu} $ satisfy the free field equations of motion,
$\partial^2 \chi_{0\mu} =0$.

More interesting are the space-space components
of metric $h_{ij}$. The equations of
motion are not diagonal for them but after a simple linear redefinition
$f_{ij} = h_{ij}-\delta_{ij} h_{00}$ they are diagonalized and have the form:
\beq{
f_{ij,\alpha}^{,\alpha} - {2\over \tau} f_{ij,0} =0
}\label{nonconf}
\eeq
This is the same equation which is satisfied by a massless minimally coupled
scalar field. The solutions with definite momenta
can be written as
\beq{
\phi_k (\tau,\vec x) = H\left( \tau - {i\over k}\right) \exp(i\vec k \vec x -
ik\tau )
}\label{sol}
\eeq
These nonconformal modes differ from the conformal modes by the presence of the
$i/k$ term.

The conventionally normalized Heisenberg field operator for quantum
fluctuations of the gravitational field is defined as $\psi_{\mu\nu} \equiv
 h_{\mu\nu}/\kappa$. This
may be expressed in the standard way in terms of creation-annihilation
operators $a_{\mu\nu}(k)$ by the decomposition:\footnote{See, e.g.,
Ref.~\cite{bdbook} for further details.}
\beq{
\psi_{\mu\nu} = \int {d^3k \over (2\pi)^3 \sqrt{2k}}
\left[\psi_k (\tau,\vec x) a_{\mu\nu} (\vec k) + \psi^* (\tau, \vec x)
a_{\mu\nu}^{\dag}(\vec k) \right]
}\label{quant}
\eeq
with $a$ and $a^{\dag}$ satisfying the commutation relations
\beq{
\left[a_{\mu\nu} (\vec k), a^{\dag}_{\mu\nu} (\vec q)\right]
=(2\pi)^3 \delta^3 (\vec k -\vec q)
\hskip1cm{\rm(not~summed~on}~\mu,\nu\rm)}\label{commut}
\eeq
All other commutators are zero.  For the nonconformal modes, the wave function
$\psi_k (\tau,\vec x)$ is given by Eq.~(\ref{sol}), while for the conformal
modes, they are plane waves $H\tau \exp(i\vec k \vec x - ik\tau ).$
Respectively, the propagator for a conformal field (for example for the field
$h_{0\alpha}$, see above) is proportional to that in flat space.
It means that the evolution of $h_{0\alpha}$
is trivial and in particular the corresponding Green's functions are
obtained from the ones for the flat spacetime by the rescaling:
\beq{
G^{dS} (x, x') = {G^{flat} (x,x') \over a(\tau) a(\tau')}
=H^2 \tau\tau' G^{flat} (x,x')
}\label{confgreen}.
\eeq

Now, propagators (two-point functions) for nonconformal fields exhibit new
features
since the solutions (\ref{sol}) blow up at small $k$ as
compared to the solution for conformal fields. In particular, this implies
that the anticommutator correlator $G_1$ is in fact not defined in the
infrared.  To be explicit, it is given by
\beq{
G_1(\tau,\tau',\vec k)={H^2\over k}\bigl[\bigl(\tau\tau'+{1\over
k^2}\bigr)\cos(k\Delta\tau)-{\Delta\tau\over k}\sin(k\Delta\tau)\bigr],
}\label{gone}
\eeq
where $\Delta\tau\equiv \tau-\tau'.$
If we calculate the Fourier transform of (\ref{gone}) then it is
logarithmically
divergent at small $k$ and undefined. However, derivatives of it are well
defined.

It might also worth mentioning that in a de~Sitter background, the Green
function for nonconformal field can be also given in a closed form in four
dimensional notations. Namely we have
\beq
G(x,x')~=~\tau\tau'\int{{d^4k}\over{(2\pi)^4}}{{\exp~i(k,x-x') }
\over {k^2}}~-~\int {{d^4k}\over {(2\pi)^4}}
{{\exp~i(k,x-x')}\over {k^4}}\label{exact}
\eeq
where, as usual, depending on the $i\epsilon$ prescription we get either
Feynman or retarded, or advanced propagator. (The derivation of (\ref{exact})
follows the lines presented in the appendix to Ref \cite{parker} where the
first terms
in the WKB expansion are considered for arbitrary background field. In a
de~Sitter background, a similar argument fixes the exact propagator.) In
particular
the Feynman propagator takes the form\cite{tw2}
\beq
G_F(x,x')~=~{1\over{4\pi^2}}\left[{{\tau\tau'}\over{(x-x')^2}}-{1\over 2}\ln
(x-x')^2\right]~.
\eeq
Note that the argument of the log is not defined, as a manifestation of the
same infrared instability. What is specific about the de~Sitter background
(and $\xi=1/6~or~0$) is that the expansion in $k^{-2}$ in Eq.~(\ref{exact})
terminates on the second term.

In fact, this infrared blowup of the Feynman propagator is the basic
observation which gave rise to the hopes that quantum corrections
to the de~Sitter metric grow with time. Indeed, the quantum propagators
are not vanishing in causally forbidden regions and may bring information on
the overall expansion in the future, characteristic of the de~Sitter solution,
revealing in this way an instability of pure classical solution. We are going
to scrutinize this suggestion.

The best way to approach the problem is to keep as close to classical
consideration
as possible. Indeed, the classical solutions are known to be stable
against perturbations \cite{lifshitz}. The crucial difference between
classical and quantum problems is that
development of classical fluctuations is governed by the retarded Green
function
$G_R$
which is free of the infrared divergencies mentioned above.
The Green
function is obtained in the standard way:
\beq{
G_R (x, x') =i \theta (\tau -\tau ') \langle \left[\psi (\tau, \vec x),
\psi (\tau', \vec x')\right] \rangle
}\label{gret1}
\eeq
where we have suppressed the various tensor indices.
Although commonly expressed in terms of a VEV, it is important to recognize
that $G_R$ for the linearized theory is a purely classical construct that may
be obtained directly from the classical equations of motion.  Accordingly, it
is completely independent of the definition of the vacuum state.

A simple
calculation gives \cite{tw3}
\beq{
G_R(x,x') = H^2[\tau\tau' G_R^{flat} (x-x') + \theta(\Delta \tau)
\theta (\Delta \tau - r)/4\pi]
}\label{gret2}
\eeq
where $r=|\vec x - \vec x'|$, $\Delta \tau = (\tau - \tau')$ and
\beq{
G_R^{flat} = \theta (\Delta\tau) \delta (r -\Delta \tau) /4\pi r
%- \delta (\Delta \tau +r )
}\label{gflat}
\eeq
is the retarded Green's function in the flat spacetime. We suppressed here
the evident tensor indices.  The second term in Eq.~(\ref{gret2}) is connected
with the broken conformal invariance.  For what follows it is essential that
it vanishes for $\Delta \tau =0$. In the mixed $(\tau, k)$ representation we
find
\beq{
G_R(\tau,\tau',\vec k)={H^2\theta(\Delta\tau)\over 2k}
\bigl[\bigl(\tau\tau'+{1\over k^2}\bigr)\sin(k\Delta\tau)-{\Delta\tau\over
k}\cos(k\Delta\tau)\bigr],
}\label{gret}
\eeq
and in the limit if $k\rightarrow 0$ the retarded Green function for a
nonconformal field is no more singular than for a conformal (or in flat space).
%{\bf Double-check the preceding formulas.}

The Heisenberg picture of quantum field theory is most close to the classical
one and we expect that the infrared problem is most easily treated within this
approach. With the initial condition $\psi_{\mu\nu}
(\tau_0,\vec x)= \psi_{\mu\nu}^{in} (\tau_0,\vec x)$, the operator equation
(\ref{first}) can be rewritten in the integral form:
\beq{
\psi_{\mu\nu}(\tau, \vec x) = \psi_{\mu\nu}^{in}(\tau,\vec x) +
{1\over \mpl} \int^\tau_{\tau_0} d\tau'\int d^3x'
G_{R\,\mu\nu}^{\alpha\beta} (\tau, \vec x;
\tau',\vec x') (H\tau')^{-2} V_{\alpha\beta} (\tau', \vec x')
}\label{integ}
\eeq
where $V_{\alpha\beta}(\tau,{\bf x})$ is the interaction term.
More specifically,
\beq{
V_{\mu\nu} = \psi^{\alpha\beta}_{,\mu} \psi_{\alpha\beta ,\nu} +
\psi^{\alpha\beta}\psi_{\alpha\beta ,\mu\nu}
-\psi_{,\alpha}\Gamma^\alpha_{\mu\nu} + 2\Gamma_{\mu\nu}^\alpha
\Gamma_{\alpha\beta}^\beta - 2\Gamma_{\mu\beta}^\alpha \Gamma^\beta_{\nu\alpha}
+2\eta_{\mu\nu} \left(
{\psi^{\alpha\beta} \Gamma_{\alpha\beta}^0 \over\tau} -
{2\psi_{0\alpha} \psi^{0\alpha} \over \tau^2} \right)
}\label{rhs}
\eeq
where $\Gamma^\alpha_{\mu\nu} = (\psi^\alpha_{\mu,\nu} +
\psi^\alpha_{\nu,\mu}-
\psi_{\mu\nu}^{,\alpha} )/2$ is the Christoffel symbol associated with the
metric $g_{\mu\nu} =\eta_{\mu\nu} + \kappa\psi_{\mu\nu}$ to first order in
$\psi_{\mu\nu}$.\footnote{We remind the reader that counterterms are
implicitly included in $V_{\mu\nu}$.} Note that  the last two terms that are
explicitly singular in $\tau$ are multiplied by combinations of fields whose
wave functions, in lowest order, vanish when $\tau \rightarrow 0.$  This
result persists in higher order, as we shall argue below.

We will be looking into the perturbative solution of (\ref{integ}). The use
of (\ref{integ}) insures that at least one propagator attached to
each vertex is a retarded one and is smooth in infrared limit.
Another important advantage of the Heisenberg representation is that
the state vectors in
Hilbert space remain constant and the evolution of the system is contained in
the time dependence of the field operators. Therefore one may study the time
evolution of the physical quantities such as, e.g., energy density as the
expectation values of the corresponding field operators over the initial
quantum state $|in\rangle$ which we take to be a no-particle, de~Sitter
invariant state. The time
evolution of this state is quite a complicated problem by itself since the
notion of particle production is ambiguous in general relatively (for
discussion see, e.g., \cite{bdbook}). In the Heisenberg picture we do not
confront this problem.

Although analytically quite complicated, it is straightforward
in principle to develop the perturbative solution of the integral equation
(\ref{integ})  (See Fig.~1a.)  Each three-point vertex carries a factor of
$1/\mpl$, so, for consistency,  we
should also include in $V_{\mu\nu}$ higher-point vertices obtained from the
expansion of Eq.~(\ref{eqmot}) as well as vertices of the same order from
higher dimensional operators implicitly included in the action
Eq.~(\ref{action}).  We will discuss that in more detail in a subsequent paper
where we will also consider in greater detail the renormalization of the
theory.   Here we concentrate only on a possible existence of singularities at
$\tau=0$.  The contribution to $\psi$ from the conformal Green's function reads
\beq{
\Delta_{conf}= {\tau\over \mpl} \int^\tau_{\tau_0} {d\tau'\over \tau'}
(\tau - \tau') \int d\Omega V(\tau', \vec x + \vec r)
}\label{dconf}
\eeq
where $|\vec r|=\tau-\tau'$ and the angular integration is made over directions
of $\vec r$.
The nonconformal part coming from (\ref{gret2}) contains two-terms, the first
of the form of Eq.~(\ref{dconf}) and the second given by
\beq{
\Delta_{nonconf}={1\over 4\pi\mpl}\int^\tau_{\tau_0} {d\tau' \over \tau'^2}
\int_0^{(\tau-\tau')} dr r^2 d\Omega V(\tau', \vec x + \vec r)
}\label{dnconf}
\eeq
Our interest is in the VEV of this expression.  Using translation invariance,
one may simply set the spatial arguments to $\vec 0$ to obtain
\beq{
\langle\Delta_{nonconf}\rangle = {1\over 3\mpl} \int^\tau_{\tau_0}
{d\tau'\over \tau'^2 } (\tau-\tau')^3 \langle V(\tau', \vec 0)\rangle
}\label{dsing}
\eeq
The generic form of the loop expansion is illustrated in Fig.~1b.  In
calculating
the VEV, one encounters quantum correlation functions of the type
\beq{
G_1(\tau,\tau',\vec x-\vec x')\equiv \langle 0, in|\{\psi_{in}(\tau,\vec x),
\psi_{in}(\tau',\vec x')\}|0, in\rangle.
}\label{anticomm}
\eeq
 These correlators are not time-ordered and replace the Feynman propagators of
the more familiar in-out formalism.  Unlike the retarded propagator or
commutator function, these are truly quantum-mechanical amplitudes that do not
vanish for spacelike separated points.  They depend upon the definition of the
vacuum state $|0,in\rangle$, about which we shall have more to say shortly.

The dimension of $V_{\mu\nu}$ is mass$^4$.  At one loop, it involves a single
graviton propagator, which is proportional to $H^2$.  It can be shown that
each additional loop brings out another factor of $H^2/\mpl^2.$ Thus there
are two powers of mass to be accounted for.  Since the only remaining scale on
which the VEV  $\langle 0,in|V_{\mu\nu}(\tau, \vec 0)|0, in\rangle$ can
depend is $\tau$, one would therefore naively expect it to behave as
\beq{
{ H^2\over\tau^2 } \biggl({H^2\over \mpl^2 }\biggr)^{L-1}
}\label{power}
\eeq
times possible powers of $\log|\tau|$.  This behavior, when inserted into the
integral equations, implies that the conformal modes are finite as
$\tau\rightarrow0$, while the nonconformal modes may diverge as a power of
$\log|\tau|.$  This would suggest that the de~Sitter background would be
unstable
to quantum fluctuations, since the fluctuations would be growing with time.
Although this is a much weaker singularity at $\tau=0$ than previously
suggested, \cite{tw1,tw2,tw3}, this may nevertheless lead to a breakdown a
perturbation theory and leave open the question of the ultimate future of the
de~Sitter metric.

It might worth emphasizing that it is not only the use of $G_R$ that softens
the infrared behaviour but the form of $V_{\mu\nu}$ as well.
With one exception, the nonconformal fields $\psi_{ij}$
enter the expression $V_{\mu\nu}$ in the form $\psi_{ij,\alpha}$, and the
derivative renders their contributions infrared finite.
The one exception is the second-to-last term in Eq.~(\ref{rhs}), which
includes $\psi_{ij}\Gamma_{ij}^0 /\tau.$  The connection $\Gamma_{ij}^0$
involves space derivatives of conformal modes $\psi_{0i}$ as well as the time
derivative of the nonconformal modes $\psi_{ij,0}$.   Despite what one might
think, $\psi_{ij,0}$ can be shown to satisfy a {conformal field equation, so
that all these terms in $\Gamma_{ij}^0$ involve an explicit factor of $\tau$
and satisfy the conformal integral equation.  Thus, this term in $V_{\mu\nu}$
is a kind of cross product of a nonconformal field with a conformal field.  It
is only the expectation value of squares of conformal fields that manifest
infrared problems, so one would not expect this to give trouble.
In conclusion, there are no infrared divergences in quantities of
interest to us here, and we may ignore for our purposes the fact that the
true vacuum state is not de~Sitter invariant.

Let us consider higher loop corrections to $\langle
V(\tau,\vec 0)\rangle.$  The two-loop corrections take the generic form
\begin{eqnarray}
F_2(\tau',\vec 0)  = {1\over\mpl^2} \int^{\tau'}_{\tau_0} {d\tau_1 \over
\tau_1^2}
\int^{\tau'}_{\tau_0} {d\tau_2 \over \tau_2^2} \int {d^3k\over{(2\pi)^3}}
{d^3q\over{(2\pi)^3}}& &\hskip-0.8cm
\partial' G_R(\tau', \tau_1, \vec k)
\partial' G_R(\tau', \tau_2, -\vec k)\nonumber\\
 &\times&\hskip-0.3cm \partial^2 G_1(\tau_1,\tau_2,\vec q)\partial^2
G_1(\tau_1,\tau_2,\vec k-\vec q)~.
\label{2loop}
\end{eqnarray}
Here we have gone over to a mixed time-momentum representation, which is more
convenient, since momentum is conserved, and power-counting is more familiar
in momentum space.  $G_R$ is, as before, the retarded Green function, while
$G_1$ is the anticommutator function\footnote{There are also two-loop
contributions from the commutator function.} given in Eq.~(\ref{anticomm}).
Each partial derivative in Eq.~(\ref{2loop}) is to be interpreted either as a
derivative with respect to time or multiplication by one power of momentum.
As mentioned previously, these remove the potential infrared problem normally
associated with the presence of $G_1.$   With regard to the momentum
integration, there will be various terms.  In a mass-independent
renormalization prescription, such as dimensional regularization with minimal
subtraction, all divergences proportional to a power of the cutoff are
completely removed by the counterterms.  Thus, we need only concern ourselves
with terms that are logarithmically divergent.  Such contributions cannot
change the power behavior of the time dependence, so that we simply have the
result that one would guess on the basis of dimensional analysis.  Thus, at
worst, $F_2\sim (\log|\tau'|)^p/\tau'^2$ for some power $p$.  When inserted
into the integral equation for the nonconformal $\psi_{ij}$, this yields a
result that behaves at worst as $(\log|\tau|)^{p+1}.$

Thus it is safe to say that quantum correction may bring only logarithmic
dependence on $\tau$.  Terms proportional to powers of $\ln|\tau|$ are not
inherently problems for perturbation theory.  Consider, for example, the
effect of a finite renormalization of the curvature, so that the scale factor
$a(t)$ might be of the form $\exp[(H+\delta H)t]$.  Expanding in powers of of
the perturbation, gives
\beq{
\exp[(H+\delta H)t]=\exp(Ht)[1+\delta H t +{(\delta H t)^2\over2}+....]
}\label{expand}
\eeq
In terms of the original conformal time $Ht\equiv-\ln(H|\tau|)$, these
corrections take the form of a power of $\ln|\tau|$.  Thus, it may be that
logarithms arising from loop corrections, while suggesting a breakdown of
perturbation theory, are merely finite renormalizations of the Hubble
constant.  To understand whether logarithmic corrections are a true
instability rather than simply a finite renormalization, one would have to show
that the breakdown is not simply due to large logs that can be summed up
(as is often done using the renormalization group).  Their physical
interpretation may be simplified by computing their contribution to a gauge
invariant quantity, such as the curvature.

Let us now outline an argument according to which these
logs may in fact severely constrained further.  To this end, we return to the
consideration of two-loop correction (\ref{2loop}) and Fig.~2. We have not
completely settled the issue of how many logs are present, but we can present
an argument that no logs arise from the quantum loop integral.  To this end, we
will show that the one-loop corrections to the effective action does not
contain a $\ln|\tau|$ term in the two-point function that enters the two-loop
calculation.  The r.h.s.\ of Eq.~(\ref{2loop}) indicates in fact two distinct
steps in the calculation.
First, we calculate the one-loop effective action which involves
acausal propagators. Then we iterate this action with the help of the retarded
(i.e., classical) propagators. Our procedure so far was to look for log
dependence of the generic
term (\ref{2loop}). We did not check that the result adds up to a general
covariant function. In fact, as we will demonstrate now, this constraint
may eliminate some infrared logs.

To this end let us consider the effective action generated by a scalar field
of mass m and parameter $\xi$ (see, e.g., chapter 6 of book \cite{bdbook}):
\beq
L_{div}~=~-(4\pi)^{-{n\over2}}\left({1\over {n-4}}+{1\over
2}(\gamma+\ln(m^2/\mu^2)
\right)\cdot a_2(x)\label{ano1}
\eeq
where $\gamma$ is Euler's constant, and
\beq
a_2(x)~=~{1\over {180}}R_{\alpha\beta\gamma\delta}R^{\alpha\beta\gamma\delta}
-{1\over {180}}R_{\alpha\beta}R^{\alpha\beta}-{1\over 6}({1\over 5}-\xi)
R_{;\alpha\alpha}+{1\over 2}({1\over 6}-\xi)^2R^2\label{ano}
.\eeq
Here $R_{\alpha\beta\gamma\delta}, R_{\alpha\beta}, R$ are the Riemann tensor,
Ricci tensor and scalar curvature, respectively.
Since quantum part of the gravitational field, $h_{\mu\nu}$, is a collection
of fields with $\xi=0$ or $\xi=1/6$ (see above for a discussion,)
Eq.~(\ref{ano})
can be directly used to evaluate the effective action generated by quantum
gravity. Note also that to regularize the expression in ultraviolet and in
infrared one introduces both dimensional regularization, $n\neq 4$, and a
nonvanishing mass, $m\neq 0$.

Eq (\ref{ano1}) is valid in arbitrary background field. For our purposes
we need to calculate $R_{\alpha\beta\gamma\delta}$ for the metric specified by
(\ref{psi}) and expand the result in $\psi_{\mu\nu}$. What is specific about
de~Sitter background is that
\beq
\delta \left(\sqrt {-g} a_2(x)\right)_{de~Sitter}~\sim~(n-4)g^{\mu\nu}
\psi_{\mu\nu} \label{var}
.\eeq
This means that in the limit $n\rightarrow 4$ only anomalous terms survive
where, by the anomalous terms, we now understand terms proportional to
$(n-4)^{-1}$ . Thus the correct result for the variation is not zero but
local terms.

This kind of technique is widely used to fix the trace anomaly (see, e.g.,
\cite{bdbook}) in arbitrary background. What is specific about the de~Sitter
background is that there are no ``normal" terms in one-loop expectation value
 of
the energy-momentum tensor $\langle T_{\mu\nu}\rangle $; the whole of $\langle
T_{\mu\nu} \rangle$ is anomalous both for $\xi =1/6$ and for $\xi = 0$.
While considering logs piecewise, without collecting terms into $R^2$ or other
invariants, we lose the property (\ref{var}) which in fact extends to second
derivative as well, as we argue next.

Indeed, what is special about the terms in the r.h.s. of eq (\ref{ano}) is
that, upon multiplication by $\sqrt {-g}$ they do not depend on the Hubble
constant $H$. Therefore
\beq
{{\delta^n}\over {(\delta H)^n}}(\sqrt{g}R^2)~=~0
\eeq
for any $n$. The same is true for all the invariants quadratic in
$R_{\alpha\beta\gamma\delta}$ entering $a_2(x)$. On the other hand, in
conformal coordinates, the variation of $H$ is equivalent to a variation
of metric proportional to the metric of de~Sitter space itself:
\beq
{{\delta}\over{\delta H}} g_{\mu\nu}~=~-{2\over H}g_{\mu\nu}\label{part}
.\eeq
Thus there exists a direction along which even finite variation of $g_{\mu\nu}$
leaves $\sqrt{-g}a_2(x)$ invariant. High symmetry of the de~Sitter solution
implies then that variations of $\sqrt{-g}a_2(x)$ are strongly constrained for
arbitrary change of $g_{\mu\nu}$ as well. In particular, the first variation
is always proportional to $g_{\mu\nu}$ itself:
\beq
\delta \left(\sqrt{-g}a_2(x)\right)_{de~Sitter}~=r(x)g^{\mu\nu}\delta
g_{\mu\nu}
\eeq
where $r(x)$ is function of coordinates. Since we know that the variation
vanishes
for the particular choice (\ref{part}) of $\delta g_{\mu\nu}$ we conclude that
in fact $r(x)=0$.

For our purposes here we need the
second variation of the effective action above.
This vanishes again for $n=4$. Indeed the
second derivative generically takes the form:
\beq
\delta^2 \left(\sqrt{-g}a_2(x)\right)_{de~Sitter}~=~N^{\alpha\beta
\gamma\delta}\delta
\delta g_{\alpha\beta} g_{\gamma\delta} \label{2der},
\eeq
where $N_{\alpha\beta\gamma\delta}$ is a tensor which in case of de~Sitter
space is constructed on $g_{\mu\nu}$ alone. Moreover it vanishes upon
multiplying by the variation of the special form (\ref{part}). It means,
\beq
N_{\alpha\beta\gamma\delta}~\sim~g_{\gamma\alpha}g_{\delta\beta}-
g_{\gamma\beta}g_{\delta\alpha}\eeq
and the second derivative (\ref{2der}) vanishes for any variation $\delta
g_{\mu\nu}$ symmetric in $\mu\nu$.  Thus we have shown that
 the second variation of the action
can produce only anomalous local terms. Such terms are to be included in
renormalization of the bare action and are not relevant to infrared logs.
So far we relied on infrared regularization by finite mass. If we put the mass
equal to zero from the very beginning, then one may argue\cite{parker2}
that one has $\ln(R)$ instead of $\ln(m)$. This form of the infrared cutoff
does not modify our conclusions. Indeed, if one calculates variations of
such an effective action, the result is again local terms.

To summarize, we have demonstrated that quantum corrections to the classical
de~Sitter solution generated by higher loops in quantum gravity can be at worst
powers of logs of the conformal time $\tau$.  Moreover calculation of these
logs is no longer a pure infrared problem.  We presented then further arguments
based on consideration of the one-loop effective action according to which
one-loop effects in de~Sitter background reduce to renormalization of the
action for $\psi_{\mu\nu}$, with no logs involved.  One needs also to include
the effects of the FP ghosts as well as higher-point vertices of the same order
in $1/\mpl$.

We would like to thank R. Woodard for interesting discussions and for
communicating his results prior to publication. We are also grateful to
A.I. Vainshtein and M.B. Voloshin for useful discussions. A. Dolgov is
grateful to the
Department of Physics of the University of Michigan for the hospitality during
the period when this work was done.  This work was supported in part by the
U.S. Department of Energy.

\vfill\eject
\centerline{Figure Caption}
\bigskip

\noindent Fig1a:  Perturbative solution of Eq.~21.  The line denoted by {\bf R}
corresponds to the retarded propagator.

\noindent Fig1b:  Loop corrections to metric fluctuations $\psi_{\mu\nu}$.

\vfill

\end{document}